\documentclass[aps,twocolumn,superscriptaddress,prd]{revtex4-1}

\usepackage{bm,graphicx,textcomp,amssymb,amsmath,dcolumn,color}

\usepackage[utf8]{inputenc}
\usepackage[T1]{fontenc}

\newcommand{\nn}{\nonumber\\}

\newcommand{\f}[1]{\mbox{\boldmath$#1$}}

\newcommand{\na}{\mbox{\boldmath$\nabla$}}
\newcommand{\bea}{\begin{eqnarray}}
\newcommand{\ea}{\end{eqnarray}}
\newcommand{\eea}{\end{eqnarray}}
\newcommand{\ord}{\,{\cal O}}

\begin{document}

\title{Dynamically assisted tunneling in the impulse regime}

\author{Christian Kohlf\"urst}

\affiliation{Helmholtz-Zentrum Dresden-Rossendorf, 
Bautzner Landstra{\ss}e 400, 01328 Dresden, Germany,}

\author{Friedemann Queisser}

\affiliation{Helmholtz-Zentrum Dresden-Rossendorf, 
Bautzner Landstra{\ss}e 400, 01328 Dresden, Germany,}

\affiliation{Institut f\"ur Theoretische Physik, 
Technische Universit\"at Dresden, 01062 Dresden, Germany,}

\author{Ralf Sch\"utzhold}

\affiliation{Helmholtz-Zentrum Dresden-Rossendorf, 
Bautzner Landstra{\ss}e 400, 01328 Dresden, Germany,}

\affiliation{Institut f\"ur Theoretische Physik, 
Technische Universit\"at Dresden, 01062 Dresden, Germany,}

\date{\today}

\begin{abstract}
We study the enhancement of tunneling through a potential barrier $V(x)$ by a 
time-dependent electric field with special emphasis on pulse-shaped vector 
potentials such as $A_x(t)=A_0/\cosh^2(\omega t)$. 
In addition to the known effects of pre-acceleration and potential 
deformation already present in the adiabatic regime, as well as energy 
mixing in analogy to the Franz-Keldysh effect in the non-adiabatic (impulse) 
regime, the pulse $A_x(t)$ can enhance tunneling by ``pushing'' part of the 
wave-function out of the rear end of the barrier. 
Besides the natural applications in condensed matter and atomic physics, 
these findings could be relevant for nuclear fusion, where pulses $A_x(t)$ 
with $\omega=1~\rm keV$ and peak field strengths of $10^{16}~\rm V/m$ 
might enhance tunneling rates significantly. 
\end{abstract}

\maketitle

\section{Introduction} 

One of the most striking differences between classical and quantum mechanics is 
the tunnel effect. 
Even though this phenomenon cannot be directly observed with the naked eye 
\cite{footnote-zeno}, it plays an important role in many areas of physics 
and across a wide variety of length scales -- including ultra-cold atoms in 
optical lattices on the micrometer scale (see, e.g., \cite{Bloch}),
and ranging down to nuclear physics in the femtometer regime \cite{G28,GN11}. 
However, although tunneling is usually taught in the first lecture course on 
quantum mechanics, our intuition and understanding, especially regarding 
time-dependent scenarios, 
is still far from complete \cite{footnote-time}.

In order to be more specific, let us consider the one-dimensional Schr\"odinger 
equation describing a particle with energy $E$ and mass $m$ incident on a 
potential barrier $V(x)$. 
Then, using the standard WKB approximation, we may derive the tunneling 
probability $\cal P$ (or, more precisely, its exponential contribution)  
and arrive at the following famous expression, which, 
as Sidney Coleman \cite{coleman} puts it, ``every child knows'' ($\hbar=1$)
\bea
\label{tunneling-exponent} 
{\mathcal P}\sim
e^{-2{\mathfrak S}_E}
=
\exp\left\{-2
\int\limits_{x^E_{\rm in}}^{x^E_{\rm out}}
dx\,\sqrt{2m[V(x)-E]}
\right\}.
\ea
Here $x^E_{\rm in}$ and $x^E_{\rm out}$ are the classical turning points 
with $V(x^E_{\rm in})=V(x^E_{\rm out})=E$ (assuming that there are only 
two of them).
Alternatively, one could employ the instanton picture where ${\mathfrak S}_E$ 
denotes the Euclidean action \cite{coleman}. 

Adding a temporal dependence $V(t,x)$, however, the situation becomes far 
more complex. 
Only in certain limiting cases, such as slowly varying or rapidly 
oscillating potentials $V(t,x)$, simple expressions similar to 
Eq.~\eqref{tunneling-exponent} can be derived via the quasi-static or 
the time-averaged potential approximation. 

In the following, we do not consider the general case $V(t,x)$, but we 
focus on the simpler (yet non-trivial) scenario of a static potential
barrier superimposed by a purely time-dependent electric field. 
In the Coulomb gauge, this case can be represented by a space-time 
dependent potential $V(t,x)=V_0(x)+xV_1'(t)$, 
but we find it more convenient to represent the electric field via 
the vector potential $A(t)$ in the temporal gauge. 

In order to understand the impact of the additional electric field
on the tunneling probability~\eqref{tunneling-exponent} of a charged
particle, let us try to separate the phases of the temporal evolution: 

First, even before hitting the barrier, the electric field can accelerate 
(or decelerate) the particle which effectively increases (or decreases) 
the energy $E$ and changes the classical turning points $x^E_{\rm in}$ 
and $x^E_{\rm out}$.
This effect already occurs in the adiabatic regime of slowly varying 
electric fields. 

In contrast, the time-dependence of $A(t)$ can induce additional phenomena
in the non-adiabatic (impulse) regime. 
Thus, as the second effect, the energy $E$ is no longer conserved due to this 
temporal variation, i.e., it can effectively shift the energy up or down. 
Assuming a harmonic oscillation $A(t)=A_0\sin(\omega t)$, for example,
this effect can be understood in the Floquet picture \cite{Grifoni}
where we get side-bands with effective energies $E\pm n\hbar\omega$ 
(in analogy to the Franz-Keldysh effect \cite{Franz,Keldysh}.) 
%
In view of the exponential decay of the wave-function inside the barrier,
this effect can typically most efficiently enhance the tunneling rate 
when the mixing occurs near the front end $x^E_{\rm in}$ of the barrier. 

Third, the electric field effectively deforms the barrier and thereby 
changes the tunneling probability~\eqref{tunneling-exponent}. 
Similar to the first contribution (pre-acceleration) and in distinction to 
the second (energy mixing), this effect already occurs in the adiabatic 
regime. 

Finally, the time-dependent electric field may enhance tunneling by 
effectively ``pushing'' part of the wave-function out of the rear end 
$x^E_{\rm out}$ of the barrier in the non-adiabatic (impulse) regime. 
(Reversing the electric field would then suppress tunneling by ``pulling''
it back into the barrier.)
In the following, we shall study this fourth displacement effect and its 
relation to the other three contributions in more detail. 

Of course, the electric field does also influence the wave packet after 
tunneling through the barrier, but this does not change the tunneling 
probability -- unless the wave packet is ``pulled'' back into the barrier again. 
This case is relevant for scenarios where the time-averaged potential 
approximation applies, but it will not play a role here. 
One should also stress that the above dissection according to the various 
stages of the temporal evolution (or the spatial regions) is not sharp 
as we are dealing with wave-packets instead of point particles.  
Due to the corresponding uncertainty, all these phenomena will be 
intertwined in general. 
Only in appropriate limiting cases, it will be possible to separate 
these effects clearly, as we shall see below. 

\section{Kramers-Henneberger Map} 

The fourth displacement effect described above can be nicely understood using 
the Kramers-Henneberger transformation \cite{H68}
which describes an exact mapping of a 
purely time-dependent electric field to the induced quiver motion $\f{\chi}(t)$. 
In order to briefly recapitulate the basic principle, let us start from the 
Schr\"odinger equation 
\bea
\label{schroedinger}
i\partial_t\psi=-\frac{[\na-iq\f{A}(t)]^2}{2m}\,\psi+V\psi
\,.
\ea
For a purely time-dependent vector potential $\f{A}(t)$, we may eliminate 
the quadratic term $q^2\f{A}^2/(2m)$ by a global phase transformation 
$\psi\to e^{i\phi(t)}\psi$ with $\dot\phi=q^2\f{A}^2/(2m)$.

If we now consider a Galilei transformation $\f{r}\to\f{r}+\f{v}t$
which changes the time derivative as $\partial_t\to\partial_t-\f{v}\cdot\na$,
we see that the additional term is equivalent to a constant vector potential 
$\f{A}$ proportional to the velocity $\f{v}$. 
This identification also works for time-dependent vector potentials 
$\f{A}(t)$ which can thus be represented by the displacement 
$\f{r}\to\f{r}+\f{\chi}(t)$ given by 
\bea
\label{quiver}
\dot{\f{\chi}}(t)=-\frac{q\f{A}(t)}{m}
\,,
\ea
which is precisely the classical solution of a point particle with mass $m$
and charge $q$ in the electric field generated by the vector potential 
$\f{A}(t)$. 
Thus, the vector potential $\f{A}(t)$ can be translated into a corresponding
displacement $\f{\chi}(t)$ of the wave-function $\psi$.

\section{Rectangular Potential} 

In order to study the four effects (pre-acceleration, energy mixing, potential
deformation, and displacement) mentioned in the Introduction, let us start with 
the extremely simple case of a rectangular (box) potential of height $V_0$ 
and length $L$ in one dimension 
\bea
\label{box}
V(x)=V_0\Theta(x)\Theta(L-x)
\,.
\ea
In the absence of the vector potential $A(t)$, we would have an incident 
wave with energy $E_{\rm in}$ on the left-hand side $x<0$ of the potential, 
plus a reflected wave with the same energy $E_{\rm in}$.
Including the vector potential $A(t)$ changes the incident wave according 
to the Kramers-Henneberger transformation described above, while the 
reflected wave will in general contain a mixture of different energies $E$.
Thus we arrive at the general ansatz 
\begin{multline}
\label{incident+reflected}
\psi(t,x<0)
=
e^{-iE_{\rm in}t+i\sqrt{2mE_{\rm in}}\,[x-\chi(t)]}+\\
+\int dE\,\psi_{\rm ref}(E)e^{-iEt-i\sqrt{2mE}\,[x-\chi(t)]}
\,.
\end{multline}
The term $\sqrt{2mE_{\rm in}}\,\chi(t)$ in the exponent of 
the first line describes the acceleration by the electric field 
before hitting the barrier.  
Similarly, we use the general ansatz for the solutions inside the barrier 
\begin{multline}
\psi(t,0<x<L)=\int dE\,e^{-iEt}
\times
\\
\left(
\psi_{\rm int}^+(E)e^{+\sqrt{2m(V_0-E)}\,[x-\chi(t)]}+
\right.
\\
\left.
+\psi_{\rm int}^-(E)e^{-\sqrt{2m(V_0-E)}\,[x-\chi(t)]}\right)
\,,
\end{multline}
as well as the transmitted solutions
\bea
\label{ansatz-transmitted}
\psi(t,x>L)
=
\int\limits dE\,\psi_{\rm tra}(E)
e^{-iEt+i\sqrt{2mE}\,[x-\chi(t)]}
\,,
\ea
where we have assumed that there is no wave incident from the right-hand 
side of the barrier $x>L$. 

Now the matching conditions for $\psi(t,x)$ and $\psi'(t,x)$
at $x=0$ and $x=L$ for all times $t$ uniquely determine 
$\psi_{\rm ref}(E)$, $\psi_{\rm int}^\pm(E)$, and 
$\psi_{\rm tra}(E)$.
While the resulting set of equations is linear and thus solvable 
via numerical discretization, for example, we shall use some analytical 
approximations to gain further inside in the following. 

\subsection{Sudden approximation}\label{sudden}

As our first example, let us consider a very strong and short pulse $A(t)$
which can be approximated by a Dirac delta function $A(t)\propto\delta(t)$. 
According to Eq.~\eqref{quiver}, this corresponds to a sudden displacement 
$\chi(t)=\Delta\chi\Theta(t)$, where we assume $L>\Delta\chi>0$. 
Considering an initially stationary solution with $E_{\rm in}$,
we find two major effects:

At the front of the barrier (around $x=0$), the displaced wave function 
is no longer an energy eigenstate, implying a mixture of energies $E$. 
The contributions corresponding to higher energies $E>E_{\rm in}$ 
can then tunnel through the barrier more easily than the initial 
stationary solution with $E_{\rm in}$.

At the rear of the barrier (around $x=L$), a part of the exponential 
tunneling tail with length $\Delta\chi$ is ``pushed'' out of the barrier 
and thus the probability density behind the barrier is exponentially 
enhanced by a factor of $e^{2\sqrt{2m(V_0-E_{\rm in})}\,\Delta\chi}$,
which could be quite large. 

\subsection{Opaque-barrier approximation}

In order to treat more realistic time-dependences $\chi(t)$, we employ the 
opaque-barrier (low-energy) approximation. 
To this end, we assume that the potential height $V_0$ is much larger than 
all other energy and frequency scales, such as $V_0\gg E_{\rm in}$. 
Furthermore, we assume that the barrier width $L$ is also much larger than 
all other length scales, such as $L\gg|\chi|$. 
As a result, the tunneling rate is strongly suppressed due to 
$\sqrt{2mV_0}L\gg1$, i.e., the barrier is very opaque (see also \cite{BL82}). 

In addition to the large quantity $\sqrt{2mV_0}L\gg1$, we only keep those terms 
where an energy $E\ll V_0$ is combined with the long length $L$ 
as well as those terms where the displacement $\chi\ll L$ is combined with 
the large barrier height $V_0$. 
To be consistent, we neglect all terms which do not contain a large quantity 
($V_0$ or $L$), such as all combinations of an energy $E\ll V_0$ with a 
displacement $\chi\ll L$.
Within this scheme, the acceleration before the barrier  
$\sqrt{2mE_{\rm in}}\,\chi(t)$ in Eq.~\eqref{incident+reflected}
is neglected.  

Using this approximation scheme, we may simplify the matching conditions 
at $x=0$ and $x=L$.
Due to the exponential suppression of the tunneling rate, we have 
$\psi_{\rm int}^+(E)\lll\psi_{\rm int}^-(E)$ such that the contribution 
from $\psi_{\rm int}^+(E)$ can be neglected at $x=0$. 
To leading order in $E/V_0$, we then find 
$\psi_{\rm ref}(E)\approx-\delta(E-E_{\rm in})$ 
which yields 
\bea
\psi_{\rm int}^-(E)
\approx
-2i\sqrt{\frac{E_{\rm in}}{V_0}}
\int\frac{dt}{2\pi}\,
e^{i(E-E_{\rm in})t-\sqrt{2mV_0}\,\chi(t)}
\,.
\ea
Consistent with the simple picture described in the previous 
Section~\ref{sudden}, this equation describes the energy mixing at 
the front of the barrier $x=0$ due to the time-dependent 
displacement $\chi(t)$.
Then we may use the remaining matching condition at $x=L$ in order 
to determine the transmitted solution 
\bea
\label{transmitted-solution}
\psi_{\rm tra}(E)
\approx
\psi_E^0
\int\frac{dt}{2\pi}\,
e^{i(E-E_{\rm in})t-\sqrt{2mV_0}[\chi(t+i{\mathfrak T})-\chi(t)]},
\ea
where $\psi_E^0=4e^{-\sqrt{2mV_0}\,L+E_{\rm in}{\mathfrak T}-i\sqrt{2mE}\,L} 
\sqrt{E_{\rm in}/V_0}$ 
collects the $\chi$-independent factors and 
${\mathfrak T}$ denotes the B\"uttiker-Landauer traversal time \cite{BL82} 
for this case (with $V_0\gg E_{\rm in}$)
\bea
\label{traversal}
{\mathfrak T}=L\sqrt{\frac{m}{2V_0}}
\,.
\ea
Note that the first two exponentials 
$e^{-\sqrt{2mV_0}\,L+E_{\rm in}{\mathfrak T}}$
in $\psi_E^0$ are the leading-order contributions of the undisturbed 
tunneling exponent $e^{-\sqrt{2m(V_0-E_{\rm in})}\,L}$.

\subsection{Dynamically assisted tunneling}

The result~\eqref{transmitted-solution} now enables us to study the enhancement
of tunneling due to the vector potential $A(t)$.
In this form~\eqref{transmitted-solution}, we directly see the invariance 
under Galilei and gauge transformations $A(t)\to A(t)+\rm const$.
Furthermore, it is interesting to note that the exponent 
$\sqrt{2mV_0}[\chi(t+i{\mathfrak T})-\chi(t)]$ is consistent with the 
change of the instanton action \cite{footnote-instanton} 
%
\bea
\label{instanton-action}
\sqrt{2mV_0}[\chi(t+i{\mathfrak T})-\chi(t)]
=
-\sqrt{\frac{2V_0}{m}}\int\limits_t^{t+i{\mathfrak T}} dt'qA(t')
,\quad
\ea
because $\sqrt{2V_0/m}$ is the undisturbed instanton velocity. 
Thus, the above Eq.~\eqref{instanton-action} is the analogue to formula~(10)
in \cite{QS}. 
However, that work \cite{QS} was mainly focused on the tunneling exponent, 
whereas the above Eq.~\eqref{transmitted-solution} does also contain the 
pre-factor [for the box potential~\eqref{box} and within the used 
approximations].  

If we derive the typical energy gain or loss from 
Eq.~\eqref{transmitted-solution} via the saddle-point method 
\bea
\Delta E
&=&
E-E_{\rm in}
\approx 
-i\sqrt{2mV_0}[\dot\chi(t+i{\mathfrak T})-\dot\chi(t)]
\nn
&=&
iq\sqrt{\frac{2V_0}{m}}[A(t+i{\mathfrak T})-A(t)]
\,,
\ea
we also find agreement with the instanton picture \cite{footnote-instanton}. 
Note that the real time $t$ corresponds to the rear $x=L$ of the barrier, 
the complex time $t+i{\mathfrak T}$ to the front $x=0$, see also 
\cite{Ivlev+Melnikov-rf}.  

As expected, the time scale distinguishing the adiabatic from the 
non-adiabatic (impulse) regime is the B\"uttiker-Landauer traversal 
time~\eqref{traversal}. 

In the adiabatic regime of slowly varying $\chi(t)$, the difference 
$\chi(t+i{\mathfrak T})-\chi(t)\approx i{\mathfrak T}\dot\chi(t)$, 
is purely imaginary. 
Thus, the main effect is an energy shift $\Delta E$ given by 
$mL\ddot\chi(t)$, i.e., the energy gained in the electric field. 
A real contribution $\sqrt{2mV_0}{\mathfrak T}^2\ddot\chi(t)/2$
to the exponent in Eq.~\eqref{transmitted-solution} arises to second 
order in ${\mathfrak T}$.
This contribution can be re-written as ${\mathfrak T}\Delta E/2$ and 
just reflects the quasi-static deformation of the potential by the 
electric field. 

In the non-adiabatic regime of rapidly changing $\chi(t)$, however,
we may get a stronger enhancement of the tunneling probability -- 
as expected from the previous considerations. 
The first term $\chi(t+i{\mathfrak T})$ in the exponent in 
Eq.~\eqref{transmitted-solution} stems from the front end $x^E_{\rm in}=0$ 
of the barrier where the imaginary shift $t \to t+i{\mathfrak T}$ to 
complex time corresponds to energy mixing. 
The second term $\chi(t)$ stems from the rear end $x^E_{\rm out}=L$ and 
reflects how the wave-function is ``pushed'' out of the barrier. 
Altogether, we obtain the following rough estimate for the enhancement 
of the tunneling probability
\bea
{\cal P}_\chi\sim{\cal P}_0
\exp\left\{\ord\left(2\sqrt{2mV_0}\Delta\chi\right)\right\}  
\,.
\ea
Note, however, that the simple result~\eqref{transmitted-solution} 
has been derived for the box potential~\eqref{box} and will not apply 
to other potential barriers $V(x)$ in general.
The discontinuities at the front $x^E_{\rm in}=0$ and the rear end 
$x^E_{\rm out}=L$ make the energy mixing and displacement mechanisms 
quite efficient --  this may change for other $V(x)$.

\section{Triangular Potential}\label{Triangular Potential} 

In order to study the difference between a very steep potential slope and a more 
gradual change, let us consider a triangular potential
\bea
\label{triangular-potential}
V(x)=\frac{x}{L}\,V_0\Theta(x)\Theta(L-x)
\,.
\ea
Of course, the ansatz for the solutions outside the barrier is the same as 
before in Eqs.~\eqref{incident+reflected} and \eqref{ansatz-transmitted}.
The solutions inside the barrier can be written in terms of displaced
Airy functions Ai and Bi 
\begin{multline}
\psi(t,0<x<L)=\int dE\,e^{-iEt+i\varphi(t)}
\times
\\
\left(
\psi_{\rm int}^{\rm A}(E)\,{\rm Ai}
\left[
\left(\frac{2mV_0}{L}\right)^{1/3}
\left(x-\chi(t)-\frac{EL}{V_0}\right)
\right] 
+
\right.
\\
\left.
+
\psi_{\rm int}^{\rm B}(E)\,{\rm Bi}
\left[
\left(\frac{2mV_0}{L}\right)^{1/3}
\left(x-\chi(t)-\frac{EL}{V_0}\right)
\right] 
\right)
\,,
\end{multline}
with the additional global phase $\dot\varphi(t)=V_0\chi(t)/L$,
since an $x$ displacement of a linear potential $V(x)\propto x$
translates into a variation of the potential height. 

As before, we apply the opaque-barrier and low-energy approximation 
$\chi(t)\ll L$ and $E\ll V_0$.
In addition, the slope $V_0/L$ of the triangle is assumed to be small 
such that $V_0\chi(t)/L\ll E$. 
Under these assumptions, products of small terms are neglected. 
%
Note that this includes the above phase $\varphi(t)$ as it scales with 
the product of the small slope $V_0/L$ and $\chi(t)$. 
However, as this phase $\varphi(t)$ does also include a time integral
over $\chi(t)$, neglecting $\varphi(t)$ poses a restriction on the relevant 
time scales, which should not be too long.
This is consistent with our previous considerations, as we are interested 
in rapid (i.e., non-adiabatic) changes.

Then, in analogy to the opaque-barrier approximation for the box potential,
we find $\psi_{\rm int}^{\rm A}(E)\ggg\psi_{\rm int}^{\rm B}(E)$. 
Using the approximations described above, we find the transmitted solution 
\bea
\label{transmitted-solution-triangle}
\psi_{\rm tra}(E)
\approx
\psi_E^0
\int\frac{dt}{2\pi}\,
e^{i(E-E_{\rm in})t+\sqrt{2mV_0}\chi(t)}
\,.
\ea
where $\psi_E^0$ now describes the undisturbed ($\chi=0$) amplitude 
for the triangle~\eqref{triangular-potential}.

Comparing Eqs.~\eqref{transmitted-solution} and  
\eqref{transmitted-solution-triangle}, we find that the term 
$\chi(t+i{\mathfrak T})$ describing the energy mixing at the front end 
is missing -- just the ``pushing out'' effect at the rear end $x=L$ 
of the barrier remains. 
The suppression of energy mixing at the front end $x_E^{\rm in}=EL/V_0$ 
of the barrier (i.e., the first classical turning point) can intuitively 
be understood by the nearly adiabatic evolution in space and time induced
by the gradual change (with a very small slope).

To complete the picture, let us consider the case of a triangular potential
turned around, where the step-like discontinuity is at the front end while 
we have the gradual change at the rear end. 
As one might have expected from the previous considerations, we now obtain 
``the other half'' of Eq.~\eqref{transmitted-solution}, i.e., the term 
$\chi(t+i{\mathfrak T})$ describing the mixing of energies at the front end 
-- while the ``pushing out'' effect at the rear end has negligible effect 
due to the nearly adiabatic evolution in space and time induced by the 
gradual change.

Depending on the temporal structure of $\chi(t)$, the energy mixing 
$\chi(t+i{\mathfrak T})$ and the ``pushing out'' contribution $\chi(t)$
can be very different. 
For example, enhancing the tunneling probability by ``pushing out'' part 
of the wave-function obviously requires $\chi>0$, while the energy mixing 
contribution $\chi(t+i{\mathfrak T})$ can also enhance tunneling if the 
electric field points in the other direction -- provided that its  
temporal structure (e.g., spectrum) contains large enough frequency 
components. 

In stationary tunneling, one cannot observe such a difference between the 
triangular potential~\eqref{triangular-potential} and its mirror image.
Due to unitarity, the tunneling probability does not depend on whether 
the wave is incident from left or right. 
The time-dependent vector potential $A(t)$, however, induces a 
non-equilibrium situation, where such a breaking of symmetry is possible. 
This phenomenon is closely related to quantum ratchets, 
see, e.g., \cite{quantum-ratchets}. 

\section{Experimental Realizations}\label{Experimental-Realizations}

Since tunneling plays a role in various areas of physics, its dynamical 
assistance could be observed in several scenarios.
However, as most of them will not correspond to a rectangular or 
shallow triangular potential, let us first briefly discuss the case of 
a general potential $V(x)$. 

\subsection{B\"uttiker-Landauer traversal time}
\label{Buttiker-Landauer}

For general potentials $V(x)$, the B\"uttiker-Landauer traversal time 
${\mathfrak T}$ is given by \cite{BL82}
\bea
\label{general-traversal}
{\mathfrak T}
=
-\frac{d{\mathfrak S}_E}{dE}
=
\int\limits_{x^E_{\rm in}}^{x^E_{\rm out}}
dx\,\sqrt{\frac{m}{2[V(x)-E]}}
\,.
\ea
This quantity plays a mani-fold role.
It describes the (imaginary) propagation time of an instanton from 
$x^E_{\rm in}$ to $x^E_{\rm out}$.
Furthermore, it measures how much the instanton action ${\mathfrak S}_E$
(which determines the tunneling exponent) decreases when increasing the 
energy $E$.
As a consequence, the B\"uttiker-Landauer traversal time ${\mathfrak T}$
provides an estimate for the frequency components $\omega$ a pulse
(or time-dependent field) should contain to facilitate a significant 
enhancement of tunneling.
If the characteristic frequency components $\omega$ are too low 
$\omega{\mathfrak T}\ll1$, the energy mixing is not sufficient to 
shift the instanton action ${\mathfrak S}_E$ enough. 
As a result, the B\"uttiker-Landauer traversal time ${\mathfrak T}$ 
can be used to separate slow (adiabatic) from fast (non-adiabatic) 
processes. 

Note, however, that this quantity ${\mathfrak T}$ does not yield 
any information about the efficiency of the energy mixing or 
``pushing out'' processes. 
For example, ${\mathfrak T}$ does not depend on whether the particle 
is incident from left or right, cf.~Sec.~\ref{Triangular Potential}. 
The B\"uttiker-Landauer traversal time ${\mathfrak T}$ corresponds 
to the change of the tunneling exponent, but does not describe the 
pre-factor in front of the exponential.
In the Floquet-picture, for example, these pre-factors are determined 
by the matrix elements between the Floquet bands, which determine the 
efficiency of the energy mixing or ``pushing out'' processes. 

It should also be stressed here that the B\"uttiker-Landauer traversal 
time ${\mathfrak T}$ is not necessarily the unique answer to the question 
of how long the particle stays inside the barrier during tunneling -- 
this is indeed a non-trivial issue 
(already at the stage of a proper definition, see also \cite{RS20,W06}). 
Instead, it is an important quantity for discriminating slow from fast 
processes, as explained above.
In this role, it provides a good first estimate of the experimental 
requirements for observing assisted tunneling.
From Eq.~\eqref{general-traversal}, we may read off the rough scaling law
${\mathfrak T}=\ord(L\sqrt{m/[V-E]})$ where $L$ is again the length of 
the barrier. 
In analogy, the instanton action ${\mathfrak S}_E$ in 
Eq.~\eqref{tunneling-exponent} can be estimated via 
${\mathfrak S}_E=\ord(L\sqrt{m[V-E]})$.
Now, since ${\mathfrak S}_E$ should not be too large in order to have 
a measurable tunneling probability, we obtain the rough order-of-magnitude 
estimate ${\mathfrak T}\sim\ord(L^2m)$.  

\subsection{Ultra-cold atoms in optical lattices}

On a comparably large length scale of order micrometer, optical lattices 
generated by standing laser beams in the optical or near-optical regime 
provide a potential landscape for ultra-cold atoms in which they can tunnel 
from one potential minimum to the next one, see, e.g., \cite{Bloch}. 
The potential barrier height can be tuned by the laser strength 
(and its detuning).  
The above estimate ${\mathfrak T}\sim\ord(L^2m)$ of the B\"uttiker-Landauer 
traversal time scales as the inverse of the recoil energy $E_{\rm R}=k^2/(2m)$
which is typically of the order of tens of kHz. 
Non-adiabatic variations should then be in the sub-millisecond regime, 
which is not beyond the experimental capabilities. 
%
An effective electric field can be generated by accelerated motion of the 
optical lattice (i.e., a real displacement in space), which can be understood 
as an inverse of the Kramers-Henneberger transformation. 
This realization allows us to study dynamically assisted tunneling for 
neutral particles such as atoms. 


\subsection{Electrons in solids}

Another prototypical example are electrons in solid-state devices with 
characteristic length scales between the nanometer and the micrometer scale. 
%
By fabricating these devices and applying gate voltages, one could even 
appropriately realize a box~\eqref{box} or 
triangular~\eqref{triangular-potential} potential. 
Due to the smaller length scales and the smaller (effective) mass of the 
electron, the B\"uttiker-Landauer traversal time is much shorter in this 
case and corresponds to frequencies ranging from the tera-Hertz to the 
infra-red regime 
which can be coupled in via real electromagnetic fields. 


\subsection{Atomic physics} 

On even smaller length scales in the nanometer or sub-nanometer regime, 
electrons tunnel from one atomic or molecular orbital to another or into 
free space (field ionization). 
%
The reduction of the characteristic length scales goes along with a further 
increase of the typical frequency scales necessary to reach the non-adiabatic
regime, which range from the optical or near-optical frequencies up to the 
keV regime for tightly bound electrons around highly charged (high-$Z$) 
nuclei. 



\subsection{Nuclear $\alpha$-decay}\label{alpha-decay}  

Tunneling on yet smaller length scales in the picometer to femtometer regime
plays an important role in nuclear physics.
As one of the first applications of quantum tunneling, Georg Gamov explained 
the Geiger-Nuttall law of nuclear $\alpha$-decay \cite{GN11} via tunneling 
of the $\alpha$-particle through the Coulomb barrier of the remaining nucleus 
\cite{G28}. 
Of course, it would be interesting to study the option for dynamically 
assisting this process, for example with the strong field generated by an 
x-ray free-electron laser (XFEL).
This topic has induced several, partly controversial, discussions in recent 
years, see, e.g., \cite{Delion,PP20}. 


Following our strategy above, let us estimate the B\"uttiker-Landauer 
traversal time for this case.
As shown in \cite{Schneider}, for example, the associated frequency 
scales are in the 100~keV to MeV regime, see also \cite{PP20}.
%
Even though the mass of the $\alpha$-particle is much larger than the 
electron mass, the extremely small length scales lead to ultra-short 
times. 
Since frequencies in the 100~keV to MeV regime are probably hard to reach 
with current or near-future XFEL facilities, one should search for 
alternatives. 
One such option for creating short enough pulses 
could be the electromagnetic field generated by an 
additional nucleus (see also \cite{Ivlev+Gudkov}) 
with an energy of order 50~MeV 
passing by at a distance of order $10^2$~femtometer. 
%
Of course, this electromagnetic field is not really spatially homogeneous 
-- but the main effects should persist, at least qualitatively. 
\section{Nuclear Fusion} 

In the following, let us study nuclear fusion in more detail
\cite{QS,lv+duan+liu,our-comment,Assisted-fusion}, which can be 
regarded as the process opposite to nuclear $\alpha$-decay 
(or, more general, nuclear fission).
Interesting examples include deuterium-tritium 
%
\bea
{}^2_1{\rm D}+{}^3_1{\rm T}\,\to\,{}^4_2{\rm He}+{}^1_0{\rm n}+17.6~{\rm MeV}
\,,
\ea
or proton-boron fusion (see, e.g., \cite{Giuffrida})
%
\bea
{}^1_1{\rm p}+{}^{11}_5{\rm B}\,\to\,3\times{}^4_2{\rm He}+8.7~{\rm MeV}
\,.
\ea
First, we focus on the most simple case of two particles with masses 
$m_1$ and $m_2$ and charges $q_1$ and $q_2$ in the initial state. 
Later we shall discuss the generalization to more complicated scenarios
such as muon-assisted fusion.  

\subsection{The model} 

Describing the two nuclei as non-relativistic point particles 
(in the low-energy regime), their dynamics is governed by the 
two-body Lagrangian
\bea
L_{12}
&=&
\frac{m_1}{2}\,\dot{\f{r}}_1^2
+
\frac{m_2}{2}\,\dot{\f{r}}_2^2
-
V(|\f{r}_1-\f{r}_2|)
+
\nn
&&
+(q_1\dot{\f{r}}_1+q_2\dot{\f{r}}_2)\cdot\f{A}(t)
\,,
\ea
where the potential $V(|\f{r}_1-\f{r}_2|)$ contains the Coulomb repulsion at 
large distances and the nuclear attraction at short distances.
The vector potential $\f{A}$ represents the field of the XFEL, which can be 
approximated by a purely time-dependent field because the XFEL wavelength is 
much larger than the characteristic length scales of our problem, 
see also~\cite{QS}. 

In center-of-mass $\f{R}=(m_1\f{r}_1+m_2\f{r}_2)/(m_1+m_2)$ 
and relative coordinates $\f{r}=\f{r}_1-\f{r}_2$, the effective 
single-body Lagrangian for the latter reads 
\bea
L=\frac{m}{2}\,\dot{\f{r}}^2-V(|\f{r}|)+q_{\rm eff}\dot{\f{r}}\cdot\f{A}(t)
\,,
\ea
with the reduced mass $1/m=1/m_1+1/m_2$ and the effective charge 
$q_{\rm eff}=(q_1m_2-q_2m_1)/(m_1+m_2)$. 
Upon quantization, we arrive at the same Schr\"odinger 
equation~\eqref{schroedinger}, but with re-scaled variables $m$  and 
$q_{\rm eff}$. 

\subsection{Scaling analysis} 

In order to identify the relevant parameters, let us first perform a scaling 
analysis of the Schr\"odinger equation~\eqref{schroedinger} which is 
facilitated by the self-similarity of the Coulomb potential. 
Neglecting the details of the nuclear attraction at short distances 
(in a low-energy approximation), the potential $V(|\f{r}|)$ is a homogeneous
function of degree $-1$, i.e., $V(|\lambda\f{r}|)=V(|\f{r}|)/\lambda$.
This allows us to cast the Schr\"odinger equation~\eqref{schroedinger} 
into a dimension-less form.
Using the initial energy $E$ in order to set the frequency and time scale, 
the length scale can be set by the outer (classical) turning point $r_E$ 
where $V(r_E)=E$. 
Then the comparison of the kinetic term $\na^2\psi/(2m)$ with the energy 
$E\psi$
yields the first dimension-less parameter 
\bea
\label{first-dimension-less-parameter}
\eta=2mEr_E^2=\frac{2m}{E}\left(\frac{q_1q_2}{4\pi\varepsilon_0}\right)^2
\,.
\ea
%
The square root of this 
parameter 
yields the undisturbed 
WKB tunneling exponent ${\cal P}\sim\exp\{-\pi\sqrt{\eta}\}$ 
in Eq.~\eqref{tunneling-exponent} and the B\"uttiker-Landauer traversal time 
$E{\mathfrak T}=\pi\sqrt{\eta}/4$ 
for this potential. 
As explained above, this WKB tunneling exponent (and thus $\eta$) 
should not become too large in order to have a measurable tunneling probability. 

The second dimension-less parameter can be constructed by incorporating the 
remaining term $q_{\rm eff}\f{A}$ in the Schr\"odinger 
equation~\eqref{schroedinger}.
Of course, in view of the dimension-less 
parameters~\eqref{first-dimension-less-parameter} and $E/\omega$, 
this construction is not unique.
Motivated by the above considerations, we choose to compare the amplitude 
of the displacement~\eqref{quiver} with $r_E$ giving 
\bea
\label{second-dimension-less-parameter}
\zeta
=
\frac{q_{\rm eff}A}{m\omega r_E}
=
\frac{q_{\rm eff}A}{mc}\,\frac{E}{\omega}\,\frac{4\pi\varepsilon_0c}{q_1q_2}
\,,
\ea
where the last ratio on the right-hand-side is the inverse of the QED 
fine-structure constant modified by the charge numbers $Z_1$ and $Z_2$
of the two nuclei. 
The first ratio is analogous to the inverse Keldysh parameter $1/\gamma$
or the laser parameter $a_0$, but now with the electron mass being replaced 
be the reduced mass of the nuclei. 
Hence, this quantity $\zeta$ will typically be smaller than unity -- 
but again it should not be too small to see a significant effect. 

The fact that the two dimension-less parameters $\eta$ and $\zeta$ in 
Eqs.~\eqref{first-dimension-less-parameter} 
and \eqref{second-dimension-less-parameter} should not be too far above 
or below unity, respectively, shows that we do not have good scale 
separation in our problem -- which makes it hard to distinguish the 
four effects (pre-acceleration, energy mixing, potential deformation, 
and displacement) mentioned in the Introduction.
Instead, they will all be intertwined. 
Still, the previous results, especially the comparison between the 
rectangular and the triangular potential, suggest that energy mixing
may be less efficient than displacement, for example, due to the 
gradual change of the Coulomb potential at the outer turning point $r_E$. 

As another lesson, we may compare deuterium-tritium with proton-boron 
fusion by means of the above scaling analysis. 
For the proton-boron system, the Coulomb strength $\propto q_1q_2$ is 
a factor of five stronger than in the deuterium-tritium case, 
while the reduced mass is roughly a factor of $3/4$ smaller. 
According to Eq.~\eqref{first-dimension-less-parameter}, the energy 
should thus be nearly a factor of twenty larger in order to achieve 
the same WKB tunneling probability. 
Note, however, that this rough estimate is based on neglecting the details 
of the nuclear attraction, i.e., the real factor will be a bit smaller 
than twenty. 
Nevertheless, as the effective charge $q_{\rm eff}$ of the proton-boron system
is approximately a factor of $5/2$ larger than in the deuterium-tritium case, 
the required vector potentials do not differ much (only by a factor of $3/2$) 
in the two cases. 
%

\subsection{Numerical results} 

\begin{figure}
\includegraphics[width=.35\textwidth]{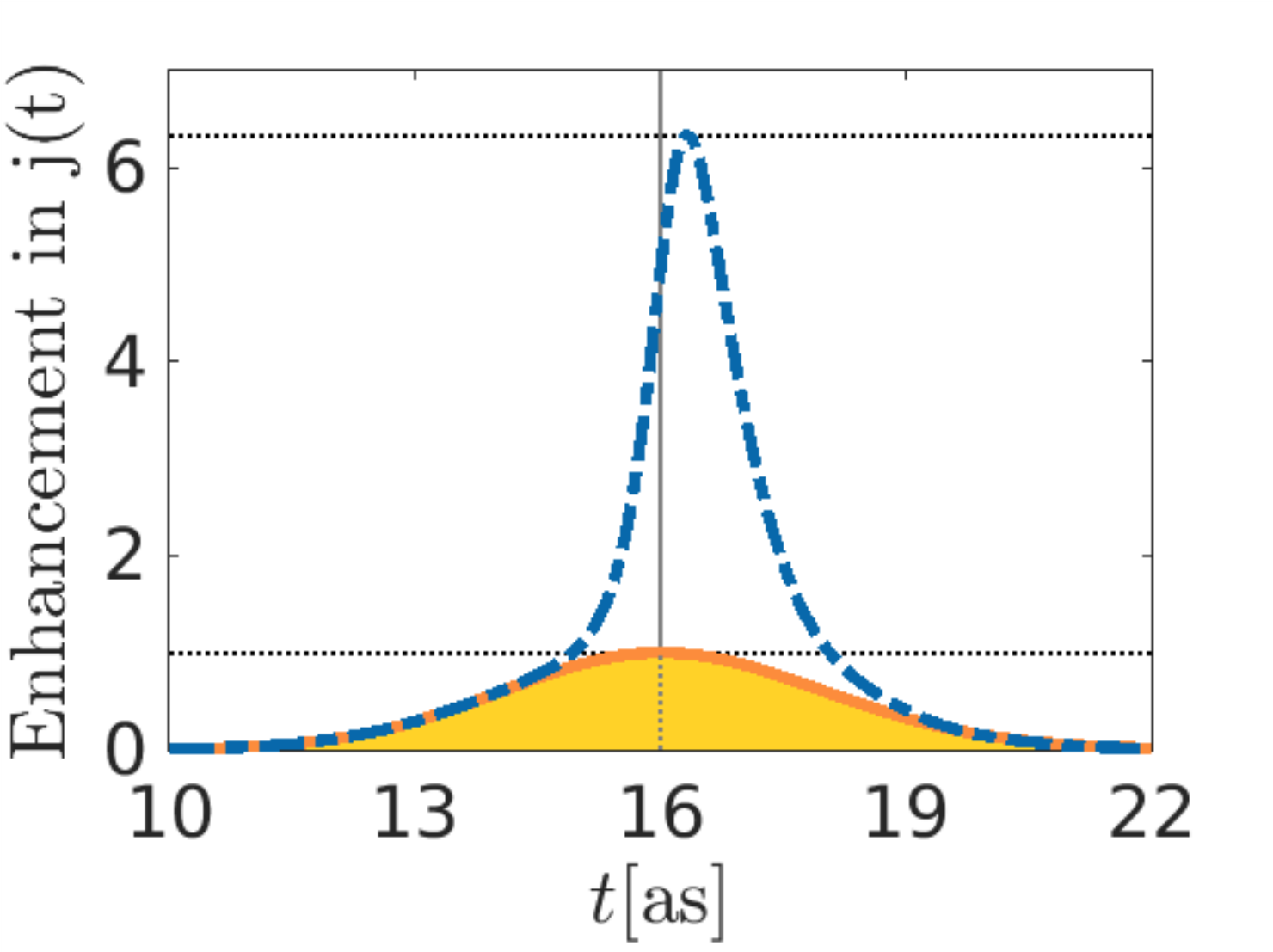} 
\includegraphics[width=.35\textwidth]{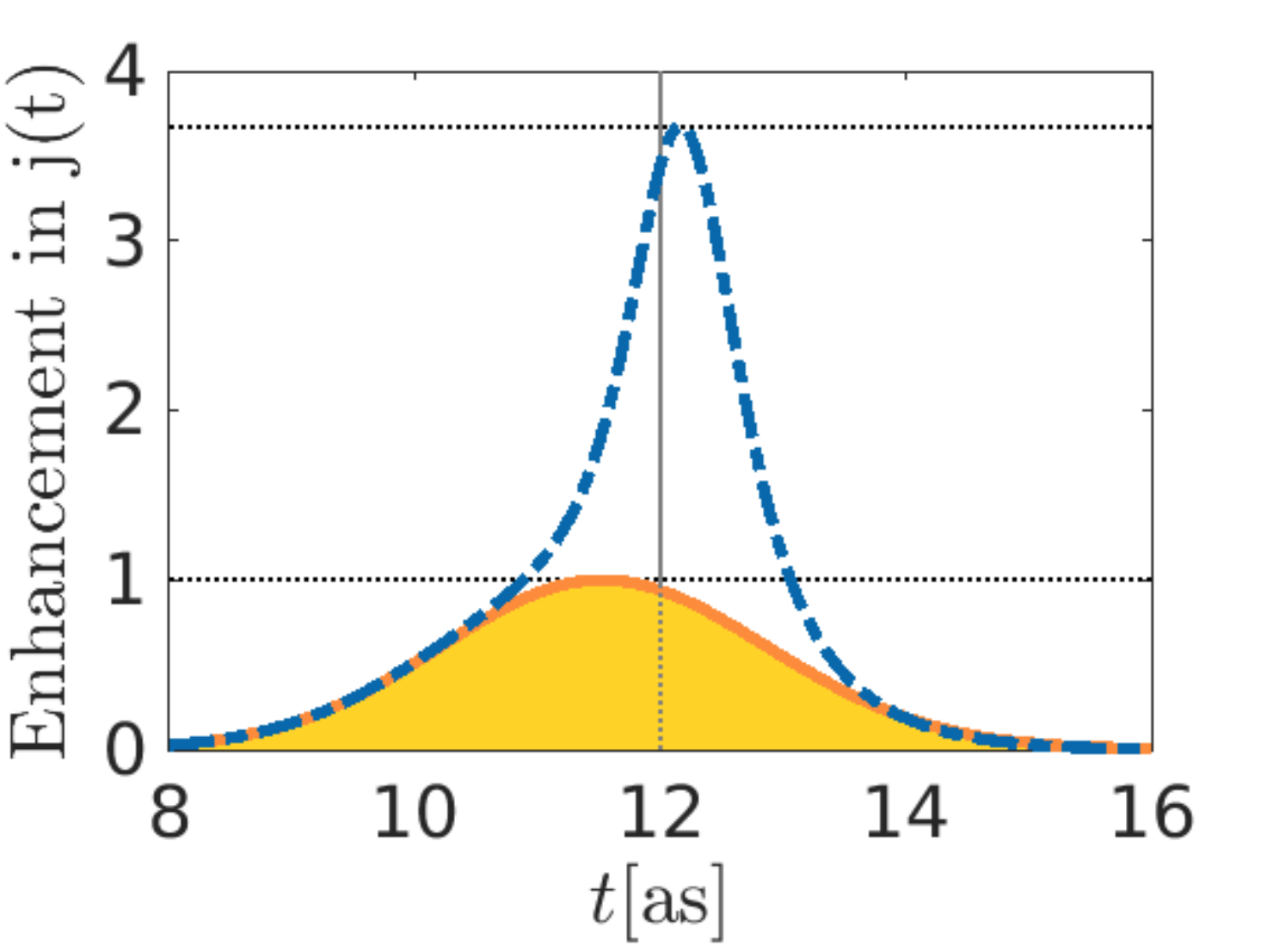} 
\includegraphics[width=.35\textwidth]{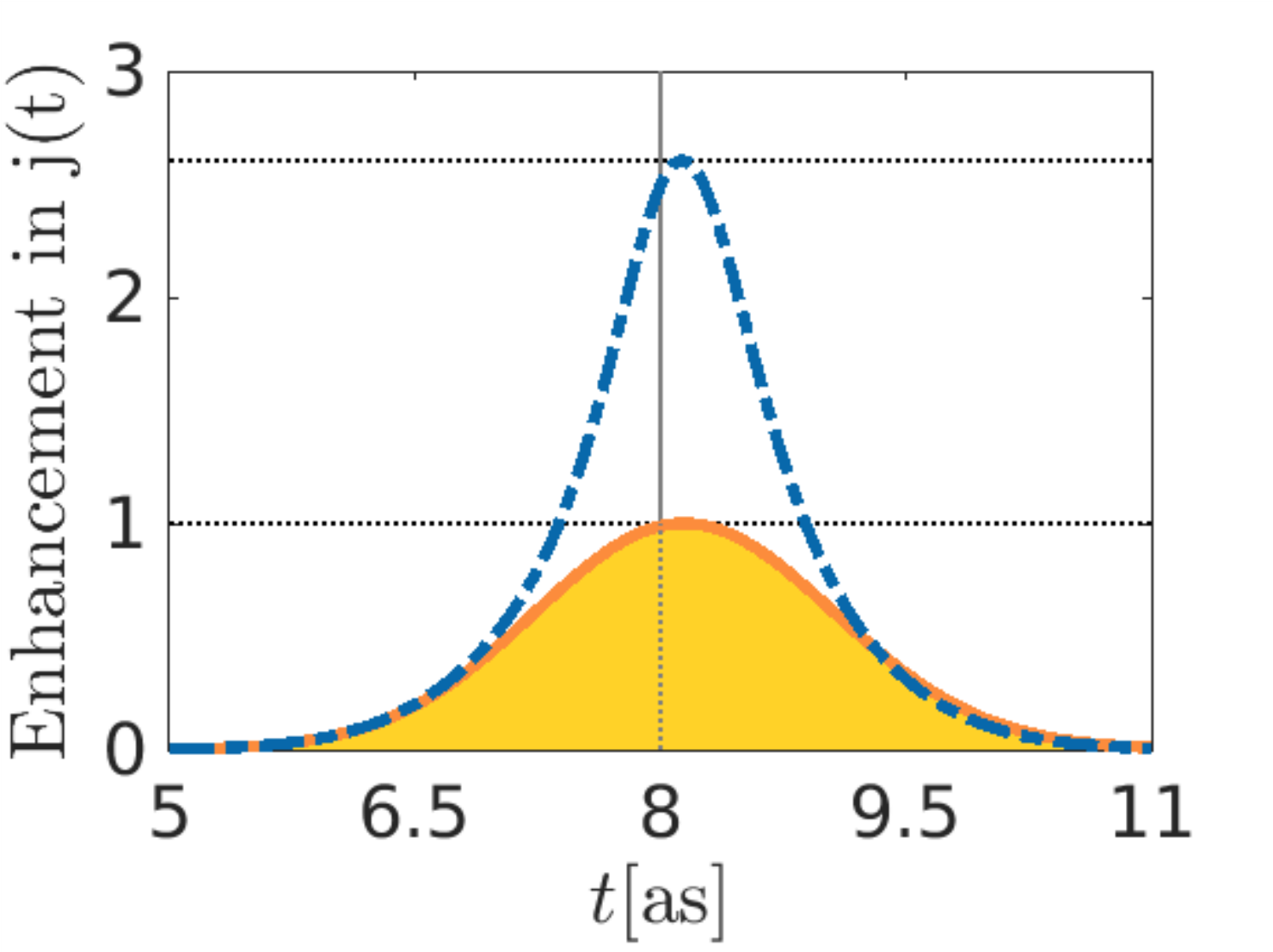} 
\caption{Enhancement of the tunneling rate for deuterium-tritium fusion with 
initial kinetic energies of 2~keV (top), 4~keV (middle), and 8~keV (bottom).
The orange curves enclosing the yellow bell-shaped regions correspond to the 
undisturbed tunneling rates $j(t)$ of the initial Gaussian wave-packets 
without the electric field while the blue dashed-dotted curves show the 
enhancement due to the pulse $A_x(t)$ with $\omega=1~\rm keV$ 
and a peak field strength of $10^{16}~\rm V/m$.}
\label{fig-d-t}
\end{figure}

\begin{figure}
\includegraphics[width=.35\textwidth]{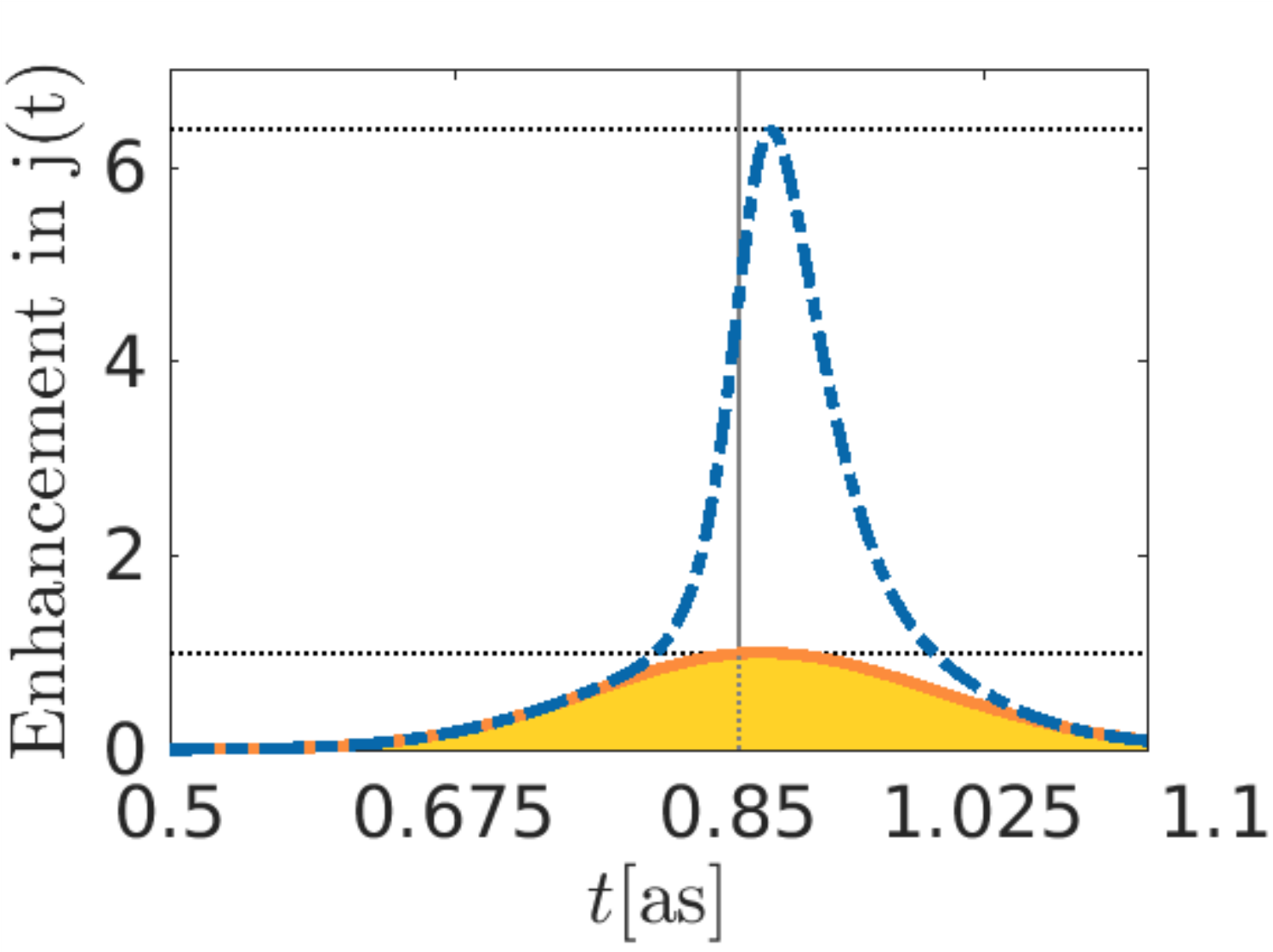} 
\includegraphics[width=.35\textwidth]{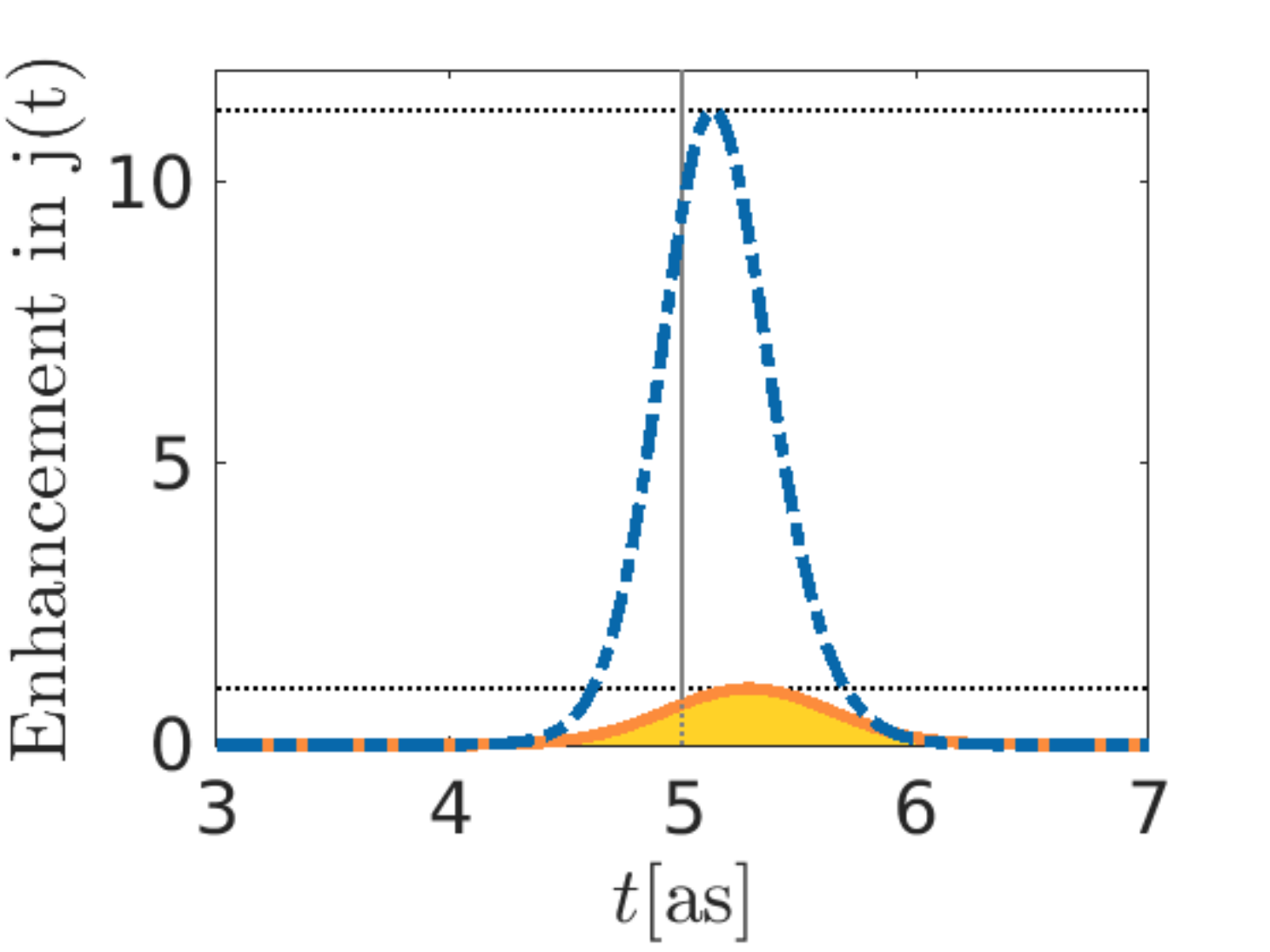} 
\includegraphics[width=.35\textwidth]{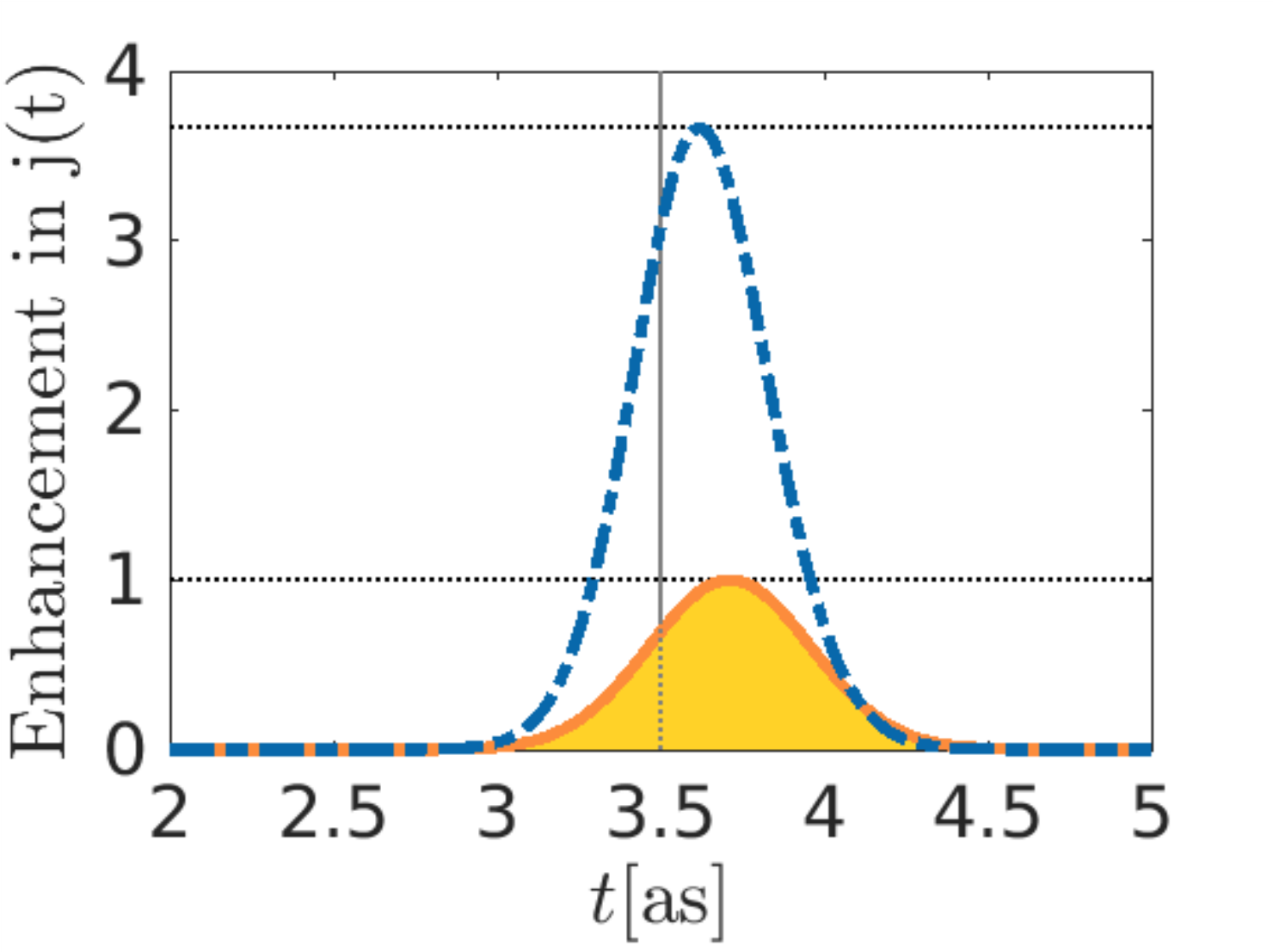} 
\caption{Enhancement of the tunneling rate for proton-Boron fusion. 
The top plot corresponds to an initial kinetic energy of 38~keV and 
a pulse with $\omega=19~\rm keV$ and $28\times10^{16}~\rm V/m$, which 
represents the scaling transformation of the scenario in Fig.~\ref{fig-d-t}
(top). 
For comparison, using the same pulse with $\omega=1~\rm keV$  
and $10^{16}~\rm V/m$ as in Fig.~\ref{fig-d-t} for deuterium-tritium fusion 
does also yield significant enhancement for initial kinetic energies of 
40~keV (middle) and 80~keV (bottom).}
\label{fig-p-b}
\end{figure}

As explained above, the analytical scaling analysis in the previous section 
does not take into account the fact that the Coulomb potential is cut off at 
nuclear distances of a few femtometer. 
In order to include this effect and to arrive at quantitative results, 
we solved the one-dimensional Schr\"odinger equation in the presence of 
the cut-off Coulomb potential $V(x)$ and the time-dependent pulse 
$A_x(t)=A_0/\cosh^2(\omega t)$ numerically for initial Gaussian wave-packets. 

Let us first discuss the case of deuterium-tritium fusion with an effective 
charge $q_{\rm eff}\approx q/5$ and reduced mass $m\approx1~\rm GeV$.
For initial kinetic energies of 2, 4, and 8~keV, the outer classical turning 
points $r_E$ determined by the Coulomb repulsion, which correspond to the 
initial turning points $x^E_{\rm in}$ in Eq.~\eqref{tunneling-exponent}, 
are $r_E\approx720~\rm fm$, 360~fm, and 180~fm, respectively. 
The inner turning point, corresponding to $x^E_{\rm out}$ in 
Eq.~\eqref{tunneling-exponent}, is determined by the nuclear attraction and 
lies around 4~fm. 

For these energies (2, 4, and 8~keV), the B\"uttiker-Landauer traversal times 
are given by ${\mathfrak T}\approx2~\rm as$, 0.7~as, and 0.25~as, respectively,
while the dimension-less parameters in Eq.~\eqref{first-dimension-less-parameter}
assume the values $\eta\approx60$, 30 and 15, respectively.
In view of $\hbar\approx0.7~\rm keV\,as$, we see that we may probe non-adiabatic 
effects with frequencies in the keV regime.  

A pulse $A_x(t)$ with $\omega=1~\rm keV$ and a peak field strength of 
$10^{16}~\rm V/m$ then corresponds to a Kramers-Henneberger displacement of 
$\Delta\chi\approx130~\rm fm$. 
%
Hence the dimension-less parameter in Eq.~\eqref{second-dimension-less-parameter}
assumes the values $\zeta\approx0.1$, 0.2, and 0.4 
for these energies (2, 4, and 8~keV). 

The enhancement of the tunneling rate (defined as the probability current 
$j(t)$ on the rear end of the barrier) is displayed in Fig.~\ref{fig-d-t}.
For all three values of the initial kinetic energy (2, 4, and 8~keV), 
we see that we obtain a significant enhancement whose relative strength 
decreases a bit with increasing energy. 
As explained above, the absence of good scale separation makes it hard to 
disentangle the four contributions (pre-acceleration, energy mixing, 
potential deformation, and displacement).
However, comparison with the results of Ref.~\cite{QS} where mainly the tunneling
exponent has been considered -- see Eq.~(10) in that work, which is the 
analogue of our formula~\eqref{instanton-action} for the box potential -- 
suggests that the energy mixing contribution is suppressed by the gradual 
change of the Coulomb potential near the outer turning point $r_E$, 
which is consistent with our results for the triangular potential. 

As motivated by the scaling analysis above, let us compare these results for 
deuterium-tritium fusion with the proton-Boron scenario.
First, in order to obtain approximately the same dimension-less parameters 
$\eta$ and $\zeta$ as in the deuterium-tritium case with $E=2~\rm keV$, 
we choose an initial kinetic energy of $E=38~\rm keV$ and a pulse with 
$\omega=19~\rm keV$ and $28\times10^{16}~\rm V/m$.
The outer classical turning point is then $r_E\approx190~\rm fm$ and the 
B\"uttiker-Landauer traversal time ${\mathfrak T}\approx0.1~\rm as$. 
Comparing the top plots in Figs.~\ref{fig-d-t} and \ref{fig-p-b}, 
we indeed find rather good agreement of the relative enhancement rates,
even though the inner turning point (a bit above 4~fm for the proton-Boron
case) does not follow the scaling transformation. 

For comparison, we also considered the impact of the same pulse 
(with $\omega=1~\rm keV$ and $10^{16}~\rm V/m$) as in the deuterium-tritium
case.
As one may observe in Fig.~\ref{fig-p-b} (middle and bottom), such a pulse 
does also yield significant enhancement rates for initial kinetic energies 
of 40 and 80~keV.
However, in view of the shorter B\"uttiker-Landauer traversal times 
${\mathfrak T}\approx0.1~\rm as$ and 0.03~as, such a pulse with 
$\omega=1~\rm keV$ may already be too slow to probe non-adiabatic effects. 

\section{Conclusions} 

We study how tunneling of a charged particle through a static potential barrier 
$V(x)$ could be dynamically assisted by an additional time-dependent electric 
field.
We identify four main mechanisms corresponding to the stages of the temporal
evolution or the spatial regions:
$(i)$ pre-acceleration before the barrier,
$(ii)$ energy mixing at its front end, 
$(iii)$ deformation of the potential barrier, and 
$(iv)$ displacement at its rear end. 
While the two effects $(i)$ and $(iii)$ are already present in the adiabatic 
regime
of slowly varying electric fields, the other two phenomena $(ii)$ and $(iv)$ 
require 
sufficiently fast -- i.e., non-adiabatic -- changes of the electric field. 

For the special cases of rectangular and triangular potential barriers, we 
we were able to disentangle the four contributions by means of approximate 
analytic solutions. 
We found that the two non-adiabatic effects $(ii)$ energy mixing and $(iv)$ 
displacement (occurring at the front and rear end, respectively) are dual to 
each other with the main difference that the former one $(ii)$ is associated 
to an analytic continuation to complex times $t\to t+i\mathfrak T$ where 
$\mathfrak T$ is the B\"uttiker-Landauer traversal time. 
This analytic continuation reflects the energy shift, which in turn changes 
the tunneling through the barrier.

The occurrence of a complex time is also related to another distinction 
between the two contributions $(ii)$ and $(iv)$. 
The displacement effect $(iv)$ strongly depends on the sign of the electric field: 
In one direction, the electric field would ``push'' parts of the wave-function 
out of the rear end of the barrier and thereby enhance tunneling -- while an 
electric field pointing in the opposite direction would ``pull'' them into 
the barrier again and thereby suppress tunneling. 
In contrast, the energy mixing effect $(ii)$ can enhance tunneling for both 
signs of the electric field -- especially deep in the non-adiabatic regime, 
where the shift of the energy induced by the time-dependence of the electric 
field becomes the most important contribution. 

Furthermore, the comparison between the rectangular and the triangular
potentials indicates that the efficiency of the two non-adiabatic effects $(ii)$
and $(iv)$ strongly depends on the shape of the potential at the turning points 
and is suppressed in the case of a gradual change.  
This finding is related to the quantum ratchet effect where tunneling in one 
direction may be favored compared to the other in non-equilibrium situations. 

The B\"uttiker-Landauer traversal time $\mathfrak T$ mentioned above is an 
important quantity in this respect because it can be used to distinguish 
slow (i.e., adiabatic) from fast (i.e., non-adiabatic) processes.
Since tunneling is a crucial effect in many areas of physics, we briefly 
discuss the observability of the considered mechanisms in several scenarios,
such as ultra-cold atoms in optical lattices, electrons in solids and in 
atoms/molecules, and nuclear $\alpha$-decay.

Finally, we turn our attention to nuclear fusion and discuss the special 
scaling properties of this process stemming from the Coulomb potential. 
As an explicit example, we study a Sauter pulse $A_x(t)=A_0/\cosh^2(\omega t)$
with $\omega=1~\rm keV$ and a peak field strength of $10^{16}~\rm V/m$ and 
find a significant enhancement of the tunneling rate for deuterium-tritium
fusion with initial energies between 2 and 8~keV as well as for proton-Boron 
fusion in the 40-80~keV energy range. 

Of course, the required field strength of $10^{16}~\rm V/m$ is quite large, 
but still well below the Schwinger critical field 
$E_S=m_e^2c^3/(q\hbar)\approx1.3\times10^{18}~\rm V/m$. 
Even though it is probably beyond the present capabilities of 
x-ray free-electron lasers (XFEL), such field strengths might be achievable 
with further technological progress, e.g., focusing the XFEL beam better.
Another interesting option could be high-harmonic focusing \cite{pukhov} 
or similar effects of light-matter interaction at ultra-high intensities. 
As a different approach, such ultra-short and ultra-strong pulses 
(though with a potentially non-negligible spatial dependence)
could be generated by charged particles (such as $\alpha$ particles) 
with sufficiently high energies (e.g., 50~keV or above) and small enough 
impact parameters (e.g., $500~\rm fm$ or below). 


\section{Outlook} 

As became evident from the previous considerations, our understanding is by far 
not complete yet and there are many ways for further progress.
For example, one could generalize the approximate analytical solutions derived
above to other potentials, such as combinations of piece-wise constant, linear, 
or even parabolic potential barriers. 
Another obvious generalization is the solution of the full 3D-Schr\"odinger 
equation, e.g., in position representation or after an expansion into spherical 
harmonics.  
Even though one would expect the main mechanisms (such as the displacement) 
to persist, there should be quantitative differences, e.g., regarding the 
pre-factors. 

This could also pave the way to study the impact of electromagnetic fields 
$\f{A}(t,\f{r})$ depending on space and time.  
Related studies of the Sauter-Schwinger effect (see, e.g., \cite{Schneider})  
show a non-trivial interplay between the spatial and the temporal 
dependence of the field. 
In this respect, it could be interesting to compare the assisted tunneling
studied in this work with the dynamically assisted Sauter-Schwinger effect 
\cite{DynamicallyAssistedSchwinger} 
which is governed by a relativistic (Dirac or Klein-Fock-Gordon) 
wave equation instead of the Schr\"odinger equation. 
%

\subsection{Muon-assisted fusion} 

\begin{figure}
\includegraphics[width=.35\textwidth]{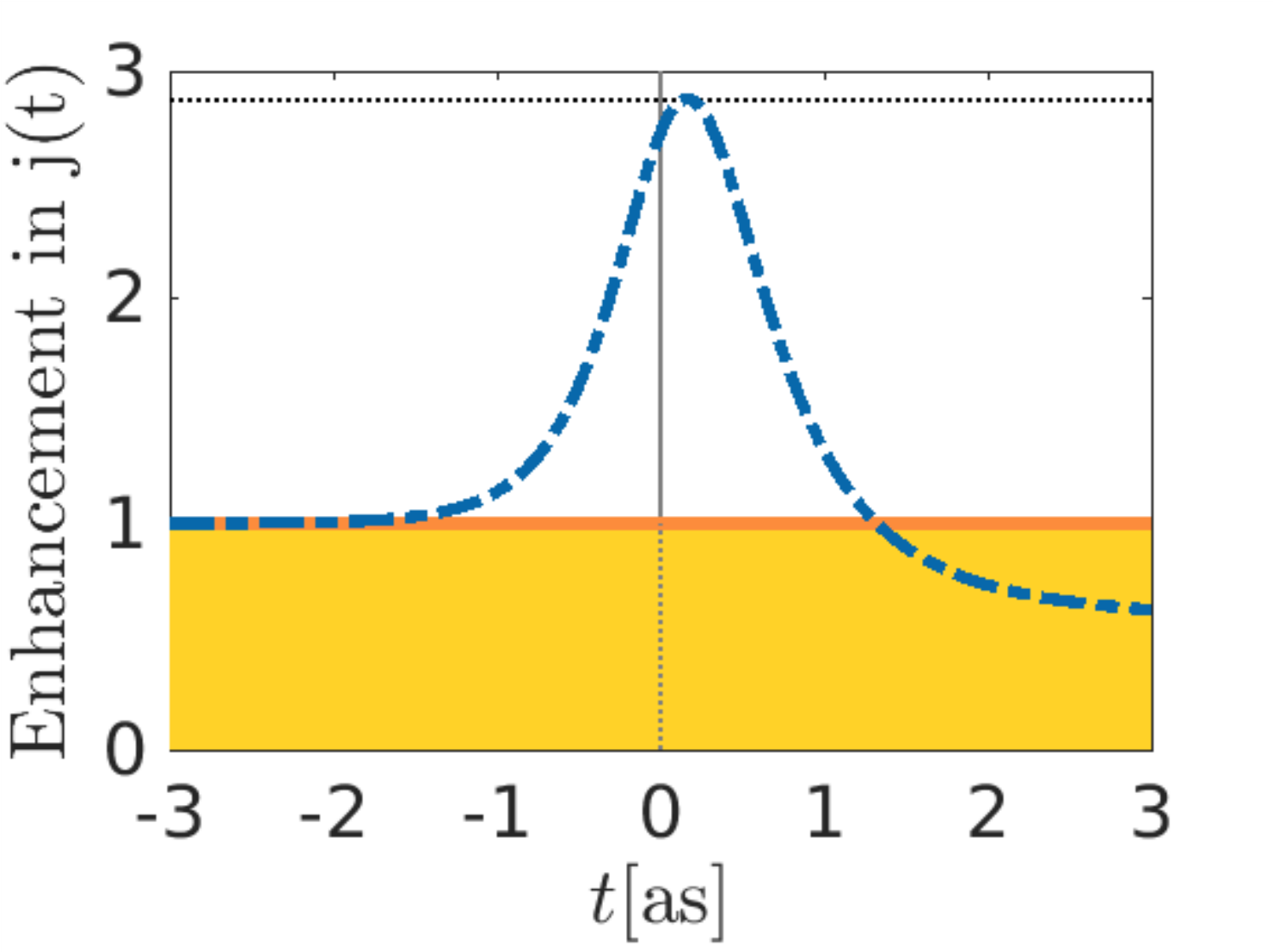} 
\caption{Enhancement of the tunneling rate starting in a bound state as a toy 
model for muon-assisted deuterium-tritium fusion, again a pulse with 
$\omega=1~\rm keV$ and $10^{16}~\rm V/m$ has been assumed.}
\label{fig-muon}
\end{figure}

As one can observe in Fig.~\ref{fig-d-t}, the relative enhancement of the 
tunneling rate does -- in stark contrast to the tunneling rate itself -- 
not depend very strongly on the initial kinetic energy, which is consistent with 
the expected behavior of the displacement mechanism, for example.
Thus, instead of starting with an asymptotically free scattering state 
(which was the scenario studied above), 
one could consider an initial bound state created by 
a dip in the Coulomb potential. 
This could be regarded as a toy model for muon-assisted nuclear fusion where 
the deuterium and tritium nuclei form a bound state with a muon, see, e.g., 
\cite{Frank,Alvarez}. 

Motivated by this toy model, we solved the Schr\"odinger equation in the 
presence of the same pulse $A_x(t)$ with $\omega=1~\rm keV$ and 
$10^{16}~\rm V/m$ as before, but now with a potential $V(x)$ admitting 
a bound state, which we used as our initial state. 
The result is plotted in Fig.~\ref{fig-muon} and shows that such a pulse 
can also induce a significant enhancement in this case. 
Even though the potential $V(x)$ is a bit different, one would expect that 
the displacement mechanism operates in a very similar way, as the vicinity 
of the inner turning point is basically unchanged.
However, the pre-acceleration mechanism should be strongly affected as it 
acts on the initial state. 

Furthermore, it should be stressed here that the case of muon-assisted 
fusion is much more complex.
For example, the electric field does not only couple to the relative 
motion of the deuterium and tritium nuclei via $q_{\rm eff}$, it also 
couples directly to the oppositely charged muon (which is much lighter).
Thus we have a real three-body problem here, which is far more involved 
-- but also offers far more interesting possibilities. 
This includes the eigen-frequencies of the three-body problem which 
partially also lie in the keV regime and thus facilitate a strong 
(and possibly resonant) coupling to the external electric field. 
For a bigger picture, one should reconsider the whole process in the 
presence of the external electromagnetic (e.g., XFEL) field, including 
the unwanted sticking of the muon to the produced $\alpha$-particles, 
which might also be affected by the external field. 


\acknowledgments 

The authors thank N.~Ahmadiniaz, R.~Sauerbrey, G.~Torgrimsson, 
and other colleagues from the HZDR for fruitful discussions.
Funded by the Deutsche Forschungsgemeinschaft 
(DFG, German Research Foundation) -- Project-IDs 398912239
and 278162697 -- SFB 1242.


\begin{thebibliography}{99}
\bibitem{footnote-zeno}
Apart from the smallness of Planck's constant $\hbar$ in everyday units, 
the laws of quantum mechanics, such as the quantum Zeno effect \cite{MS77}, 
prevent us from directly observing of the process of quantum tunneling.
%
However, sometimes the evanescent tails in more general wave propagation
phenomena are considered as analogues to tunneling.
%
These could of course be observed with the naked eye, for example in 
water waves.
\bibitem{footnote-time}
As one example, let us mention the recent controversy regarding the 
dynamical assistance of tunneling in nuclear $\alpha$-decays, 
see, e.g., \cite{PP20} and Sec.~\ref{alpha-decay}.   
%
Another example is the ongoing discussion about the tunneling time, 
see, e.g., \cite{RS20,W06} and Sec.~\ref{Buttiker-Landauer}.
\bibitem{footnote-instanton}
More details regarding the applicability of the instanton picture to 
dynamically assisted tunneling will be postponed to a forthcoming 
publication.
\end{thebibliography}
\end{document}